\documentclass[aps,prl,twocolumn,superscriptaddress,10pt]{revtex4-2}

\usepackage{bm}
\usepackage{soul, color, xcolor}
\usepackage{natbib}
\usepackage{graphicx}
\usepackage{amsmath}

\begin{document}

\title{Inelastic Neutron Scattering for Direct Detection of Chiral Phonons}

\author{Tingting Wang}
\affiliation{%
 Phonon Engineering Research Center of Jiangsu Province, Ministry of Education Key Laboratory of NSLSCS, Center for Quantum Transport and Thermal Energy Science, Institute of Physics Frontiers and Interdisciplinary Sciences, School of Physics and Technology, Nanjing Normal University, Nanjing 210023, China
}

\author{Jun Zhou}
\affiliation{%
 Phonon Engineering Research Center of Jiangsu Province, Ministry of Education Key Laboratory of NSLSCS, Center for Quantum Transport and Thermal Energy Science, Institute of Physics Frontiers and Interdisciplinary Sciences, School of Physics and Technology, Nanjing Normal University, Nanjing 210023, China
}

\author{Qingyong Ren}\email[]{renqy@ihep.ac.cn}
\affiliation{%
Institute of High Energy Physics, Chinese Academy of Sciences, Beijing 100049, China
}
\affiliation{%
Spallation Neutron Source Science Center, Dongguan 523803, China
}
\affiliation{%
Guangdong Provincial Key Laboratory of Extreme Conditions, Dongguan, 523803, China
}

\author{Lifa Zhang}\email[]{phyzlf@njnu.edu.cn}
\affiliation{%
 Phonon Engineering Research Center of Jiangsu Province, Ministry of Education Key Laboratory of NSLSCS, Center for Quantum Transport and Thermal Energy Science, Institute of Physics Frontiers and Interdisciplinary Sciences, School of Physics and Technology, Nanjing Normal University, Nanjing 210023, China
}

\begin{abstract}
Chiral phonons have attracted significant attention due to their potential applications in spintronics, superconductivity, and advanced materials, but their detection has predominantly relied on indirect photon-involved processes. Here, we propose inelastic neutron scattering (INS) as a direct and versatile approach for chiral phonon detection over a broad momentum-energy space. Leveraging INS sensitivity to phonon eigenmodes, we clearly distinguish linear, elliptical, and chiral phonons and determine phonon handedness through angle-resolved measurements. Using right-handed tellurium (Te) as a model system, we identify characteristic INS fingerprints that clearly separate chiral from linear phonons. Moreover, we show that INS can directly access phonon magnetic moments and the effective magnetic fields generated by chiral phonons, as evidenced by the pronounced mode splitting in CeF$_3$. Collectively, these results position INS as a powerful platform for comprehensive investigations of chiral-phonon dynamics and their associated quantum phenomena.
\end{abstract}

\maketitle

\textit{Introduction.- }
Chiral phonons\textendash collective lattice vibrations carrying angular momentum\textendash were proposed as an extension of chirality from fermions to bosons in condensed matter physics\cite{zhang2015chiral,zhang2014angular}. This discovery has sparked widespread interest, leading to the exploration of chiral phonons in various systems and their potential applications\cite{wang2024chiral,li2021topological,li2024utilizing}, such as regulating angular momentum in spin systems\cite{mashkovich2021terahertz,kaindl2023circling,tauchert2022polarized,sasaki2021magnetization,maehrlein2018dissecting,kim2023chiral}, enabling a novel phonomagnetic effect\cite{luo2023large,zhang2023gate,juraschek2019orbital,juraschek2022giant}, providing new pathways for high-temperature superconductivity\cite{gao2023chiral}, and identifying physical-chemical differences in biomolecules\cite{choi2022chiral,kim2022spectroscopy}. However, experimental detection and verification of chiral phonons remain challenging\cite{wang2024chiral,bourgeois2025strategy}, primarily because these spinless excitations, with large nuclear inertia, cannot be directly manipulated by external fields, and commonly appear at Brillouin-zone boundaries inaccessible by traditional optical techniques.

Currently, the main experimental methods to detect chiral phonons are based on indirect, photon-involved processes, including transient infrared (IR) spectroscopy\cite{zhu2018observation,luo2023large}, circularly polarized Raman scattering\cite{ishito2023truly,wu2023fluctuation,oishi2024selective}, and resonant inelastic X-ray scattering (RIXS)\cite{ueda2023chiral}. While these methods advanced our understanding, each has intrinsic limitations. Transient IR spectroscopy is restricted to phonons at the valley-selective K point, and Raman scattering is limited to the $\Gamma$ point of Brillouin center. Although RIXS accesses phonons at general momenta, it remains confined to systems with well-defined pseudo-angular momentum (PAM) \cite{tatsumi2018conservation,er2024circular}. Recently proposed electron scattering techniques provide a more direct probe of phonon modes, but still face PAM constraints and additional challenges under magnetic fields, limiting broader application\cite{pan2023vibrational,bourgeois2025strategy}. Therefore, a more direct and versatile detection approach is urgently needed.

To address these challenges, we propose a novel approach using inelastic neutron scattering (INS) to directly measure chiral phonons\cite{perring2009mapping,arnold2014mantid,parker2014use,ren2023extreme,squires1996introduction}. The primary advantage of INS is that it directly interacts with phonons and can couple to structures of any symmetry without being restricted by the definition of PAM. Secondly, neutron scattering pattern respond to atomic displacements, whereas Raman and IR responses are due to changes in electronic properties (polarizability and dipole moment). Therefore, while Raman and IR are often considered complementary forms of vibrational spectroscopy, INS is considered as a more powerful tool for vibrational spectroscopy, capable of measuring all phonon branches in a material. And this capability is crucial for investigating magnon-phonon coupling effects\cite{wu2023fluctuation,wang2017observation,bao2020evidence}.

In this letter, we develop a theoretical framework for identifying chiral phonons through INS and illustrate its core principles using a hexagonal lattice under simplified experimental conditions. This framework explicitly differentiates linear, elliptical, and chiral phonons, highlighting their distinct signatures. We then propose two methods capable of resolving the handedness of chiral phonons. Next, applying our approach to right-handed chiral tellurium (Te), we demonstrate how INS can clearly distinguish chiral from linear and elliptical phonons. Building on these insights, we further utilize INS to investigate effective magnetic fields induced by chiral phonons in CeF$_3$. Our results thus provide a robust experimental route for broadly probing chiral phonon, paving new ways to explore chiral phonon phenomena in diverse functional materials.

\textit{Methods.- }
To quantify the INS response, one begins from the double-differential neutron scattering cross section, which, under the one-phonon approximation, is directly related to the dynamic structure factor. This approach yields the following general expression for the INS intensity of a phonon transition\cite{parker2006neutron,parker2014use}: 
\begin{equation}
I_i\propto \sigma (\textbf{Q}\cdot \textbf{U}_i)^2exp(-Q^2U_{Tot}^2).   
\end{equation}
Here, \textbf{Q} is scattering vector, $\sigma$ is the atom-specific cross-section, $U_i$ is the phonon eigenmode amplitude. The exponential term in equation (1) is a Debye-Waller factor, $U_{Tot}$ is the total root mean-square atomic displacement of all the atoms, and its magnitude is in part determined by thermal motion. At temperatures below 30 K, the Debye-Waller factor becomes negligible \cite{parker2006neutron}. The origin of this formula and its details can be found in Supplementary Material, Section I. In this manner, the measured INS intensity provides a direct probe of phonon eigenvectors and their polarization states, including the possibility of identifying chiral phonon modes.

Chiral phonons, featuring rotational atomic motion, have not yet been directly detected experimentally, as the crucial angle-resolved detection strategy has rarely been pursued. Achieving this requires maintaining coupling to the same phonon mode while rotating the scattering angle. The INS process must satisfy momentum and energy conservation simultaneously:
\begin{align}
\textbf{k}_i - \textbf{k}_f &= \textbf{Q} = \textbf{G} + \textbf{q},\\
E_i - E_f &= \hbar \omega=\frac{\hbar^2}{2m_n}(k_i^2-k_f^2).
\end{align}
Here, $\textbf{k}_i$ and $\textbf{k}_f$ are the incident and scattered neutron wave vectors, $\textbf{G}$ is a reciprocal lattice vector, $\textbf{q}$ is the phonon wave vector within the first Brillouin zone, and $\hbar \omega$ is the phonon energy. To uncover chiral behavior, one needs to vary the direction of $\textbf{G}$ without altering $\textbf{q}$ and $\hbar \omega$. Once this limitation is taken into account, chiral phonons should, in principle, be readily detectable for the following reasons.

We consider a chiral material and focus on its $\Gamma-A$ high-symmetry path in reciprocal space. To directly probe such modes, one may select $\textbf{Q}$ wave vectors along the line from $(Q_x,Q_y,Q_z=-0.5)$ to $(Q_x,Q_y,Q_z=0.5)$ in reciprocal lattice units, corresponding to transverse phonon modes with $q$ along $A-\Gamma-A'$ and with atomic vibrations in the ab plane. By defining $Q_x^2+Q_y^2=B^2$ (where $B$ is the radial distance in reciprocal space), we can write $\textbf{G}=[Q_x,Q_y,0] = B[cos\theta,sin\theta,0]$. In an ideal scenario, where INS can measure scattering intensities at arbitrarily large wave vectors $\textbf{Q}$, the detection direction can be continuously rotated to explore the phonon trajectories at the chosen $q$ point. This strategy thus ensures direct coupling to the desired phonon modes, enabling the detection of chiral phonons through neutron scattering.

\begin{figure}[htp]
    \centering
    \includegraphics[width=\columnwidth]{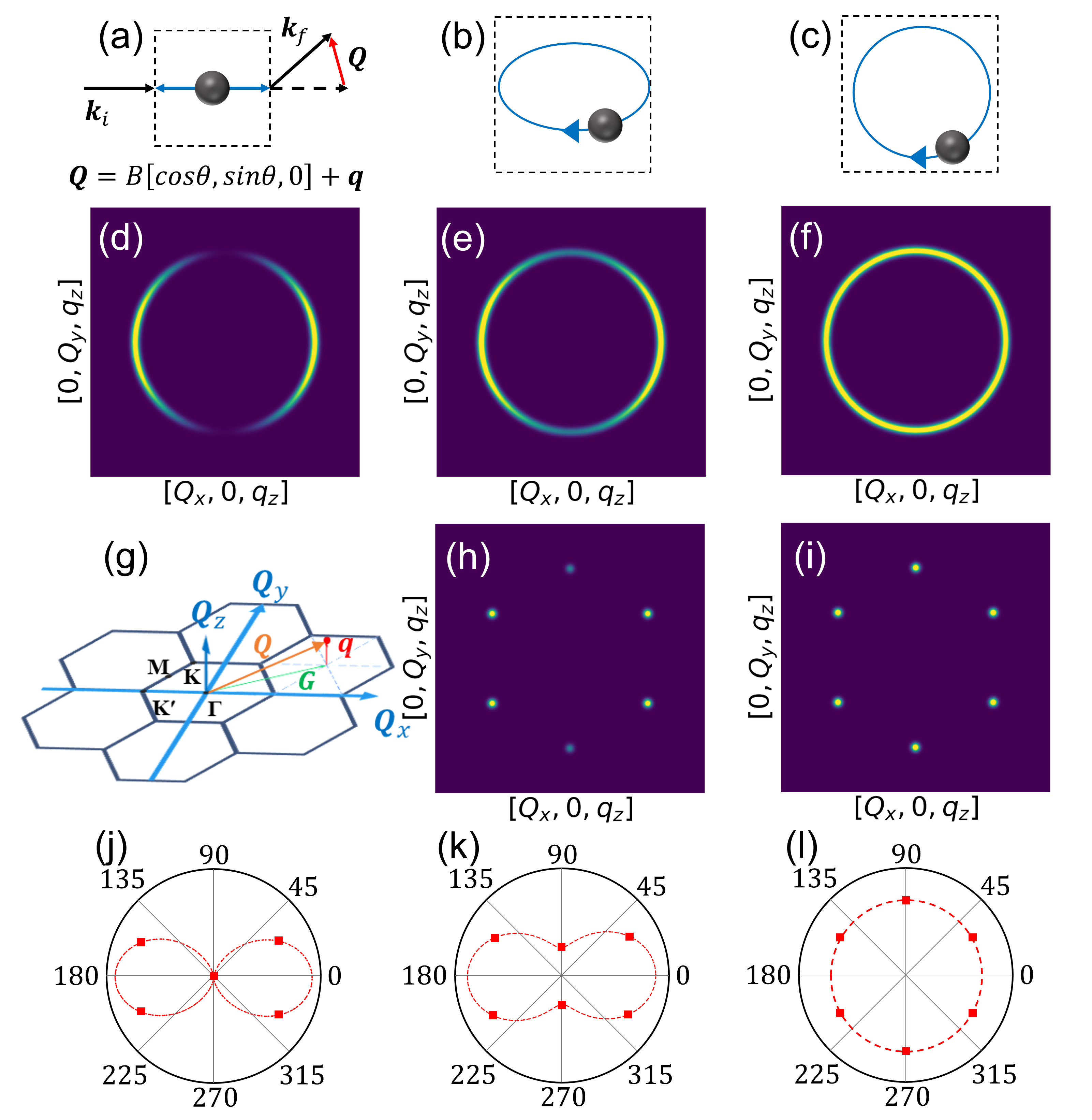}
    \caption{Schematics of (a) linear, (b) elliptical, (c) chiral phonons; and (d-f) corresponding INS intensity patterns. (g) Reciprocal lattice schematic illustrating detection of phonons with wavevector $\textbf{q}$. (h-i) INS intensity patterns for elliptical, and chiral phonons using six scattering wavevectors. (j-l) Corresponding polar intensity plots.}
    \label{fig:1}
\end{figure}

%
The atomic displacements of phonon modes measured using the above neutron scattering geometry can be expressed as a time-dependent vector: $\textbf{u}(t) = [u_x cos(\omega t), u_y sin(\omega t + \delta), 0]$, where $\omega$ is the phonon frequency. The parameters $(u_x, u_y, \delta)$ define the polarization state, with $u_x$ and $u_y$ representing the amplitudes along the $x$ and $y$ directions, respectively, and $\delta$ denoting the relative phase shift. Linear polarization is obtained when the motion collapses to a straight line, which occurs if either \(u_x=0\) or \(u_y=0\), or, equivalently, if \(u_x=u_y\neq 0\) with a phase offset \(\delta=\pm\pi/2\) (Fig. \ref{fig:1}(a)). Elliptical polarization arises when $|u_x|\ne|u_y|\ne 0$ and $\delta \ne \pm \frac{\pi}{2}$, causing the atomic displacement to trace an elliptical path in the plane, as shown in Fig. \ref{fig:1}(b). Circular polarization, is achieved when the amplitude components have equal magnitudes ($|u_x|=|u_y|=u_0$) and the phase parameter $\delta=0$ or $\pi$, resulting in the atomic displacement tracing a circular trajectory, as shown in Fig. \ref{fig:1}(c).

Typically, INS measurements are time-integrated, with the observed intensity proportional to the time-averaged quantity $\langle|\textbf{Q}\cdot\textbf{u}(t)|^2\rangle_t$. Thus, the measured intensities for linear, elliptical, and chiral phonons under different $\textbf{Q}$ are:
\begin{align}
    I_{linear}(\textbf{Q}) &\propto (B u_x \cos\theta)^2, \\
    I_{elliptical}(\textbf{Q}) &\propto B^2 \left( u_x^2 \cos^2\theta + u_y^2 \sin^2\theta \right. \nonumber \\
    &\qquad \left. - 2u_x u_y \cos\theta \sin\theta \sin\delta \right), \\
    I_{circular}(\textbf{Q}) &\propto B^2 u_0^2.
\end{align}
The details can be found in Supplementary Material, Section II(a). Consequently, linear phonons exhibit periodic angular dependence with minimum intensity dropping to zero (Fig. \ref{fig:1}(d)), while elliptical phonons retain a nonzero minimum due to orthogonal displacement components (Fig. \ref{fig:1}(e)). Chiral phonons yield a constant, angle-independent intensity-evidence of their isotropic rotational symmetry (Fig.~\ref{fig:1}(f))-after temporal integration over one oscillation period. To visualize these distributions more clearly, one can plot their angular dependence in a polar coordinate representation. The resulting polar intensity plots (Fig. \ref{fig:1}(j-l), dashed lines) distinctly reveal the “8-shaped” shape for linear polarization, an “open 8-shaped” for elliptical polarization, and a perfect circle for circular polarization, confirming their fundamental differences.

Performing continuous INS measurements over numerous scattering angles demands extensive data acquisition and significant experimental time. To overcome this practical limitation, we propose a simplified experimental design requiring measurements at only six discrete scattering angles, which suffices to clearly distinguish phonon polarization characteristics. Fig. \ref{fig:1}(g) illustrates this simplified scheme in reciprocal space, where phonons with wavevector $\textbf{q}$ are probed via neutron scattering wavevectors $\textbf{Q}$ at six adjacent Brillouin zones-a straightforward and experimentally accessible approach \cite{arnold2014mantid, hayden1991high}. Calculated INS intensity distributions in reciprocal space for linear, elliptical, and chiral phonons at these discrete scattering angles are shown in Fig. \ref{fig:1}(h-i), with brighter colors corresponding to higher intensities. Polar intensity plots-obtained by extracting the scattering angles and intensities at these six measurement points and representing them in a polar coordinate format (Fig. \ref{fig:1}(j-l), where the six points are marked as squares)-clearly distinguish the anisotropic distributions of linear and elliptical polarizations from the isotropic pattern of chiral phonons.

\begin{figure*}[htp]
    \flushleft
    \includegraphics[width=18cm]{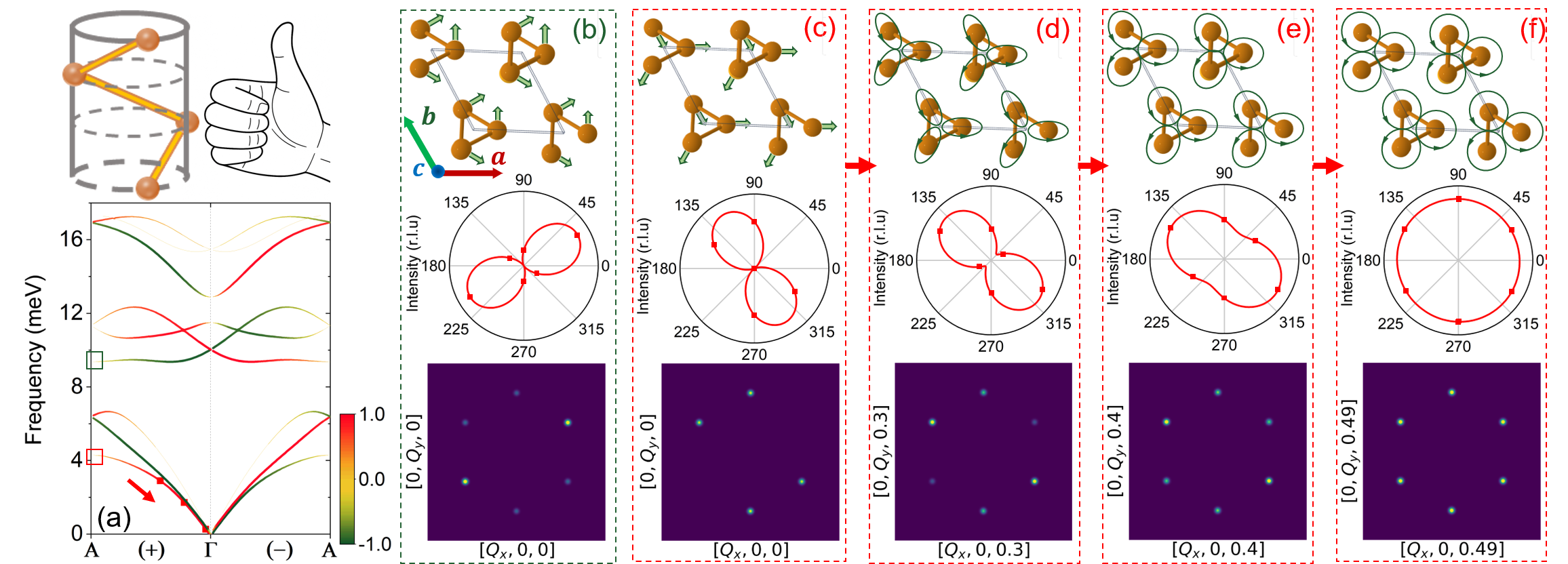}
    \caption{Theoretical INS detection of phonons in right-handed Te. (a) Structure and calculated phonon dispersion with color representing the sign of phonon angular momentum (red: positive, green: negative). (b-c) Atomic vibrations and corresponding polar INS patterns for modes marked by colored boxes in (a). (d)-(f) Evolution of atomic vibrations and polar INS patterns along the $A-\Gamma$ path.}
    \label{fig:2}
\end{figure*}

Using intensity data at six discrete scattering angles, we can reconstruct continuous polar intensity plots (Fig. \ref{fig:1}(j-l), dashed lines). This angular dependence of INS intensity naturally arises from the squared vector-dot-product form of the dynamical structure factor, which inherently allows decomposition into constant and second-harmonic trigonometric terms. Thus, the angular variation of intensity can be quantitatively described by fitting the discrete intensity data to a simplified functional form:
\begin{equation}
I(\theta) = A + C\cos[2(\theta - \phi)] + D\sin[2(\theta - \phi)],
\end{equation}
where $\phi$ denotes the intrinsic phase determined by atomic vibrational motions and reflects the orientation of the polarization relative to the crystallographic axes. This fitting approach provides robust criteria to distinguish polarization types quantitatively: linear polarization exhibits periodic intensity minima approaching zero; elliptical polarization shows nonzero minima; and circular polarization yields nearly constant intensity, independent of scattering angle.

The approach described above, which employs time-integrated INS, identifies chiral phonons through their unique isotropic scattering signatures but does not distinguish their handedness explicitly. In principle, improving the temporal resolution of INS measurements can enable explicit determination of phonon handedness by directly observing the instantaneous atomic rotational motion of chiral phonons. However, current neutron instrumentation cannot yet achieve such ultrafast measurements due to limitations in neutron flux and detector technology. A slightly relaxed alternative is indirect dichroic INS intensity measurement, which involves circularly polarized optical excitation generating nonequilibrium phonon occupations sensitive to chirality \cite{zhu2018observation,mak2012control,pan2023vibrational,pan2024strain}. Although this approach somewhat reduces the stringent requirement on temporal resolution, it remains experimentally challenging with current instrumentation. Therefore, we refrain from discussing detailed derivations and experimental considerations here, directing readers instead to Supplementary Material, Section II(b).
Having established methods to distinguish phonon polarization states and chirality using INS, we now highlight another unique capability of INS\textendash its ability to directly probe phonon magnetic moments induced by effective magnetic fields \cite{ren2021phonon,saparov2022lattice,luo2023large,juraschek2022giant,xiong2022effective}. Although phonons are electrically neutral quasiparticles, an external magnetic field couples indirectly through Lorentz forces acting on charged ions, thus modifying phonon dispersion and eigenmodes. To describe this coupling, we adopt the spin-phonon interaction Hamiltonian, widely validated by previous theoretical and experimental studies\cite{luo2023large,zhang2014angular,xiong2022effective,wang2022chiral,kariyado2015manipulation,zhang2010topological}.The magnetic-field-induced modification to the dynamical matrix, retaining linear terms in the field strength $\mathbf{H}$, is given by:
\begin{equation}
D(\mathbf{H})=D^{(0)}+i\sum_{\gamma}D^{(1)}_{\alpha \beta \gamma}H_{\gamma}, 
\end{equation}
where the coupling tensor $D^{(1)}_{\alpha \beta \gamma}$ is antisymmetric with respect to indices $\alpha$ and $\beta$ (see Supplementary Material, Sec.\,III, Eq.\,(s24)). For a doubly degenerate phonon mode at frequency $\omega$, solving the eigenvalue equation within its subspace reveals two chiral branches whose energies shift oppositely under an external magnetic field:
\begin{equation}
\omega_{H}^{\pm}=\sqrt{\omega^2\pm KH}\approx \omega\pm \frac{K}{2\omega}H,  
\end{equation}
where $K$ is the spin–phonon coupling constant. The ``$+$'' solution ($\omega_H^{+}$) corresponds to the mode with counter-clockwise atomic rotations—right-handed under our crystallographic convention—and hardens (shifts upward) for a positive field applied along $+c$. Conversely, the ``$-$'' solution ($\omega_H^{-}$) corresponds to the clockwise, left-handed mode and softens (shifts downward). Reversing the field direction interchanges these frequency shifts. The corresponding eigenvectors become:
\begin{equation}
u^\pm\approx\frac{1}{\sqrt{2}}(e_a\mp i e_b),
\end{equation}
clearly reflecting their distinct chirality. The opposite frequency shifts of $\omega_H^{\pm}$ thus provide an unambiguous spectroscopic signature to identify phonon handedness in INS experiments: observation of the upward-shifted branch directly labels it as right-handed, and the downward-shifted branch as left-handed, according to the above convention. Such chiral phonons inherently carry magnetic moments, a property that has attracted considerable attention for potential applications in ultrafast magnetism, energy-efficient spintronics, and the exploration of nonequilibrium states of matter with broken time-reversal symmetry. The unique advantages of INS, including immunity to external magnetic interference and the capability to fully resolve phonon dispersion, make it particularly suited to experimentally investigate these phonon-induced magnetic effects. For detailed derivations, see Supplementary Material, Section III.

\textit{Detection of phonon Modes in Tellurium.- }
Chiral materials inherently support a coexistence of linear, elliptical, and chiral phonon modes, making them ideal platforms for investigating distinct INS responses. Therefore, we employ first-principles calculations to simulate the phonon eigenmodes of right-handed chiral tellurium (Te), a material with well-established chiral phonon behavior \cite{zhang2023weyl,spirito2024lattice,chen2022chiral,xiong2022effective}. Computational details, structural information, and the procedure for removing complex neutron scattering phase factors—which are omitted here for clarity but experimentally accessible through symmetry-based group-theoretical analysis—are provided in Supplementary Material, Section IV. Fig. \ref{fig:2}(a) shows the structure and calculated phonon dispersion of right-handed chiral Te along the high-symmetry path $A-\Gamma-A'$, where the color represents the sign of each phonon’s angular momentum: red for positive (right-handed) and blue for negative (left-handed). At the A point the phonon mode is linearly polarized; along the $A-\Gamma$ path its polarization evolves continuously, becoming a chiral phonon at the $\Gamma$ point.

\begin{figure*}[htp] \flushleft \includegraphics[width=18cm]{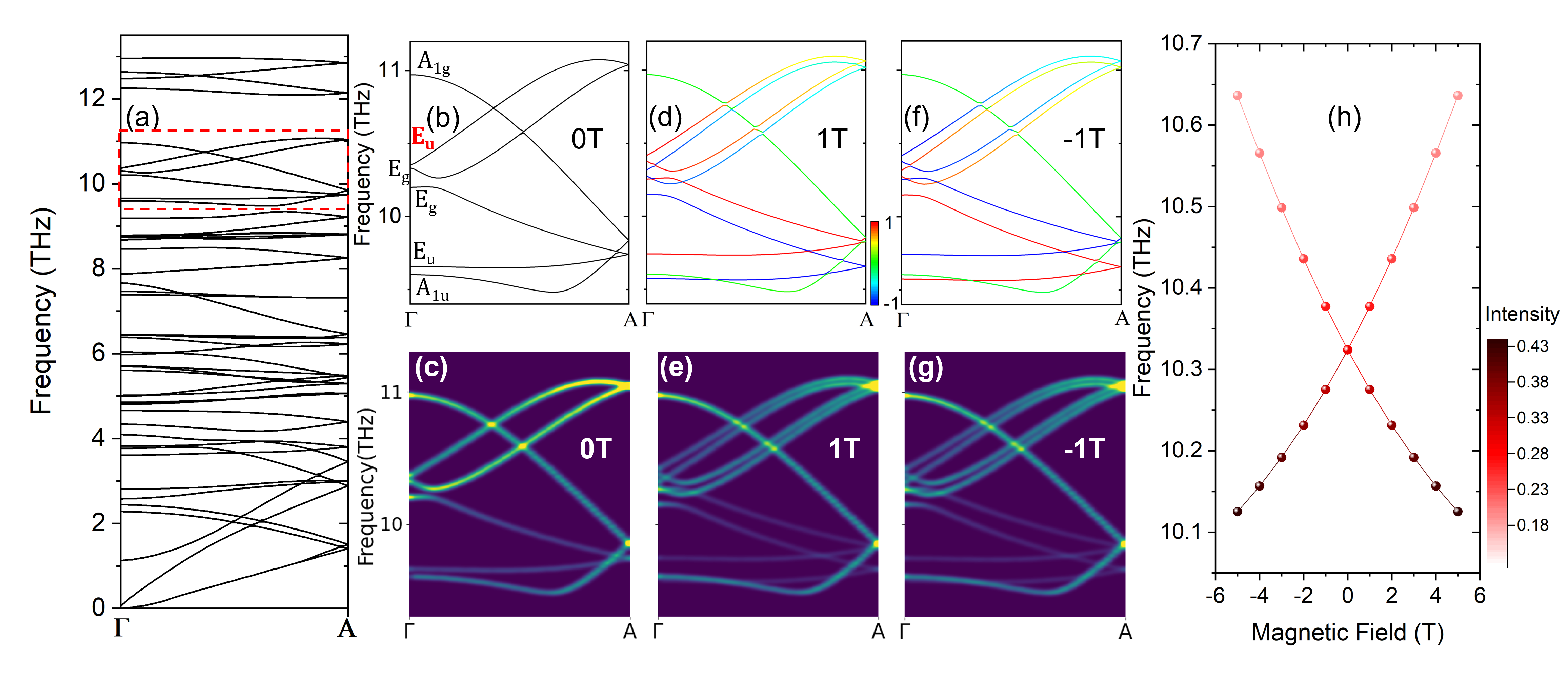} \caption{Magnetic-field-induced chiral phonon splitting in CeF$_3$. (a) Phonon dispersion along the $\Gamma-A$ path. Selected phonon branches are highlighted. (b-c) Phonon dispersion and INS intensities along the scattering vector $G=[0,0,1]$ without magnetic field. (d-g) Phonon angular momentum and corresponding INS intensities under effective magnetic fields of $H_z=\pm1$ T. (h) Splitting of the $E_u$ mode at the $\Gamma$ point under varying magnetic fields, alongside the corresponding INS intensities.} \label{fig:3} \end{figure*}

To illustrate these phonon characteristics clearly, we select representative modes marked by colored boxes in Fig. \ref{fig:2}(a). Their corresponding atomic trajectories and INS patterns at the A point (Fig. \ref{fig:2}(b-c))\textendash displayed with matching box colors\textendash exhibit the typical “8-shaped” distribution of linearly polarized phonons.
Fig. \ref{fig:2}(d-f) illustrates the evolution of phonon polarization along the red-arrow direction on the $A-\Gamma$ path, where 50 discrete q-points were sampled. As indicated by the red dot markers in Fig. \ref{fig:2}(a), the 30th, 40th, and 49th q-points capture the transition from anisotropic 8-shaped patterns at the A point to nearly isotropic circular intensity distributions near the $\Gamma$ point. These results confirm the effectiveness and sensitivity of INS in identifying distinct polarization characteristics and the dynamic evolution of chiral phonon modes.

\textit{Direct Probing of Chiral Phonon Magnetic Moments in CeF$_3$.- }
Having verified that INS can resolve the full evolution from linear to chiral phonons in right-handed Te, we next demonstrate the INS signatures of magnetic-field-induced phonon splitting in CeF$_3$. Fig. \ref{fig:3}(a) displays the phonon dispersion along the $\Gamma-A$ path, obtained from first-principles calculations. The shaded red rectangle marks the frequency window enlarged in Fig. \ref{fig:3}(b), which contains the doubly-degenerate $E_u$ branch previously shown to carry sizeable phonon angular momentum\,\cite{luo2023large}. The corresponding INS intensities along the $G=[0,0,1]$ direction without magnetic field are presented in Fig. \ref{fig:3}(c), with brighter colors indicating stronger intensities. Here, to facilitate direct comparison with potential experimental observations, we explicitly include the complex neutron scattering phase factors in the calculated INS intensities. When an effective magnetic field of $\pm1$,T is applied along the $z$-axis, the calculated phonon spectra and angular momenta are given in Figs.\ref{fig:3}(d) and (f), where red (blue) denotes right-handed (left-handed) phonon angular momentum. The companion INS intensity maps for the same scattering geometry appear in Figs.\ref{fig:3}(e) and (g). To visualise the field dependence more directly, Fig.\ref{fig:3}(h) plots the magnetic-field-induced splitting of the $E_u$ doublet at $\Gamma$; each data point is coloured by its calculated INS intensity.

In zero field the $E_u$ and $E_g$ branches are doubly degenerate and form a single dispersion. Introducing a magnetic field lifts this degeneracy, produces opposite chiralities for $\pm H_z$, and shifts the two $E$ components symmetrically in frequency. Reversing the field flips the phonon chirality, consistent with earlier theoretical predictions\cite{xiong2022effective,tang2024exciton,mustafa2025origin}. The associated INS maps show that, the splittings for $+H_z$ and $-H_z$ are similar in magnitude: because the dynamical structure factor scales as $1/\omega$, the lower (higher) branch gains (loses) intensity as $|H_z|$ increases, yielding an anti-correlated intensity contrast. This combined signature-frequency splitting together with intensity redistribution-constitutes a distinctive spectroscopic fingerprint of the phonon magnetic moment. Detailed computational parameters and convergence tests are reported in Supplementary Material, Section V.

\textit{Conclusion.- }
In summary, we proposed and validated a novel theoretical approach utilizing inelastic neutron scattering (INS) for direct detection and characterization of chiral phonons. This method, free from the symmetry and pseudo-angular momentum (PAM) constraints of existing spectroscopic techniques, clearly differentiates phonon polarization states through angle-resolved measurements. Using the prototypical chiral material tellurium (Te), we demonstrated distinct INS signatures differentiating chiral phonons from linear ones. Furthermore, we leveraged the unique capability of INS to directly probe phonon magnetic moments and effective magnetic fields induced by chiral phonons, exemplified by significant phonon splitting observed in CeF$_3$. Our results significantly advance the experimental toolkit available for studying chiral phonon phenomena and open new avenues for exploring their roles in spintronics, superconductivity, and advanced functional materials.

\begin{acknowledgments}
This work was supported by National Key Research and Development Program of China (No. 2023YFA1407001) and by Department of Science and Technology of Jiangsu Province (No. BK20220032). Q. R. acknowledges support from the National Key Research and Development Program of China (No. 2024YFE0110005) and the National Natural Science Foundation of China (No. 12474024). T. W. acknowledges support from the Postgraduate Research and Practice Innovation Program of Jiangsu Province (No. KYCX25\_1934).
\end{acknowledgments}

\bibliography{file}
\end{document}